\documentclass[runningheads]{llncs}

\usepackage[utf8]{inputenc}
\usepackage{url}
\usepackage{algorithm}
\usepackage[noend]{algpseudocode}
\usepackage{amsmath}
\usepackage{amssymb}
\usepackage{stmaryrd}
\usepackage{color}
\usepackage{todonotes}
\usepackage{url}
\usepackage{hyperref}
\usepackage{booktabs}

\makeatletter                           
\def\BState{\State\hskip-\ALG@thistlm}
\makeatother

\usepackage{graphicx}
\begin{document}

\title{Visualizing the Template of a Chaotic Attractor} 

\titlerunning{Visualizing the Template of a Chaotic Attractor}
%
\author{Maya \textsc{Olszewski}\inst{1}\orcidID{0000-0001-9926-058X} \and
Jeff \textsc{Meder}\inst{1}\orcidID{0000-0002-2360-8487} \and
Emmanuel \textsc{Kieffer}\inst{2}\orcidID{0000-0002-5530-8577} \and
Raphaël \textsc{Bleuse}\inst{1}\orcidID{0000-0002-6728-2132} \and
Martin \textsc{Rosalie}\inst{2}\orcidID{0000-0003-3676-120X} \and
Gr\'egoire \textsc{Danoy}\inst{1}\orcidID{0000-0001-9419-4210} \and
Pascal \textsc{Bouvry}\inst{1,2}\orcidID{0000-0001-9338-2834}
}
\authorrunning{M. \textsc{Olszewski} et al.}
%
\institute{FSTC/CSC-ILIAS, University of Luxembourg, 6, Avenue de la Fonte, Esch-sur-Alzette, 4364, Luxembourg
\email{\{maya.olszewski.001, jeff.meder.001\}@student.uni.lu}
\email{\{raphael.bleuse, gregoire.danoy, pascal.bouvry\}@uni.lu}
\and 
SnT, University of Luxembourg, 6, Avenue de la Fonte, Esch-sur-Alzette, 4364, Luxembourg
\email{\{emmanuel.kieffer, martin.rosalie\}@uni.lu}
}

\maketitle              
\begin{abstract}
Chaotic attractors are solutions of deterministic processes, of which the topology can be described by templates. 
We consider templates of chaotic attractors bounded by a genus--1 torus described by a linking matrix.
This article introduces a novel and unique tool to validate a linking matrix, to optimize the compactness of the corresponding template and to draw this template. 
The article provides a detailed description of the different validation steps and the extraction of an order of crossings from the linking matrix leading to a template of minimal height. 
Finally, the drawing process of the template corresponding to the matrix is saved in a Scalable Vector Graphics (SVG) file.

\end{abstract}

\section{Introduction}
Resulting of theoretical studies on chaos attractors, applications including chaotic dynamics can be found in a multitude of domains. Their range goes from computer science \cite{Rosalie_2018}, through classical sciences with physical networks \cite{Larger_2015}, biology and genetics \cite{Suzuki_2016} and chemistry with chaotic dynamics in chemical reactions \cite{Budroni_2017}, all the way to electronics and chaos in electronic devices \cite{Kumar_2017} and even environmental studies on population evolution \cite{Beninc__2015}.

Birman \& Williams \cite{birman1983knotted} introduce templates as knot-holder to describe the topological structure of chaotic attractors. The notion of linking matrices to describe chaotic attractors with integers has been first introduced by Mindlin \textit{et al.} in 1990~\cite{Mindlin_1990}. The matrix contains the number of torsions and permutations occurring along the flow of an attractor. The template is a ribbon graph combined with a layering graph.
In 1998, Gilmore wrote an extensive survey on the research on chaotic dynamical systems over the past decade~\cite{Gilmore_1998}, in which one can see various drawings of templates. 
In his paper, he provides the summary of the topological analysis from dynamical system to template.

The subject of chaotic dynamics studies are promising and on-going. But it clearly misses matrices validation and drawing tools. The research community would benefit from an efficient application that verifies the validity of matrices and draws their corresponding template. The novel tool presented in this paper is publicly available online at \url{https://gitlab.uni.lu/pcog/cate}, and aims to fill this gap.

This paper is structured as follows. In section~\ref{sec:problem description} we give an introduction to the problem. Section~\ref{sec:related work} provides a state-of-the-art analysis in the field of chaotic attractors, focusing on their validation and visualization. In section~\ref{sec:linkin matrix and template of CA}, we first outline our approach to determine the validity of a linking matrix. Secondly, we describe the procedure to get the minimal height of a template and its visualization. In section~\ref{sec:results}, we present the experimental work and the results in order to validate our proposed approach.  Finally, we conclude and outline some directions for future work in section~\ref{sec:conclusion}.

\section{Problem Description}
\label{sec:problem description}

\begin{figure}[tb]
\centering
\begin{tabular}{ccc}
\resizebox{!}{11em}{\input{malasoma_01_attra_section}} &
\qquad &
\quad \includegraphics[height=10.6em]{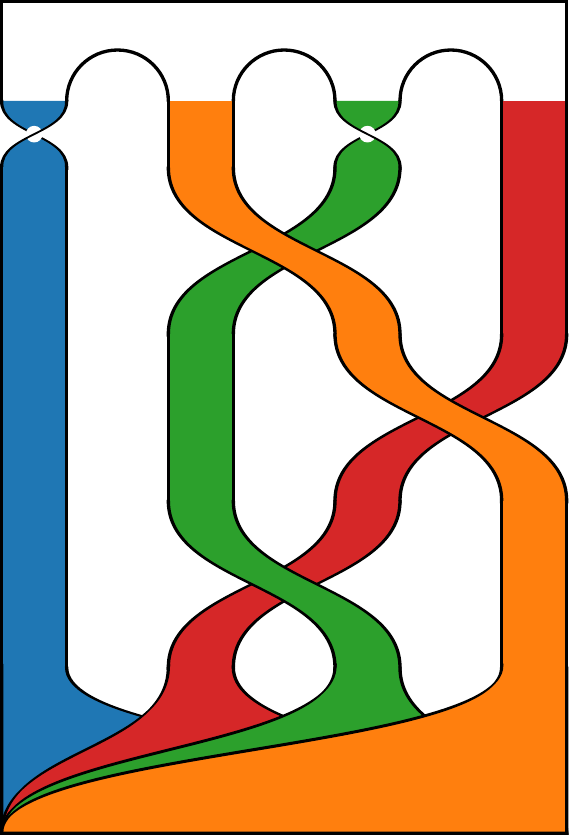}
\put(-64,89){\scriptsize 1}
\put(-45,89){\scriptsize 2}
\put(-26,89){\scriptsize 3}
\put(-6,89){\scriptsize 4}
\\
\scriptsize (a) Chaotic attractor & &
\scriptsize (b) Template
\end{tabular}
\caption{A representation of a template of a chaotic attractor solution to the Malasoma system \eqref{eq:malasoma} for $\alpha =2.027$. (a) Chaotic attractor with the Poincaré section (see \cite{Rosalie_2015} for the definition of this section named $S_a$). 
(b) Template of the chaotic attractor from the Poincaré section. 
}
\label{fig:matrix1}
\end{figure}

A chaotic attractor is a solution of a dynamic deterministic process that is very sensitive to its initial conditions. 
The solution will converge to the same global shape (the attractor), independently of the starting position in the basin of attraction.
Malasoma \cite{Malasoma_2000} proposed a simple differential equations system
\begin{equation}
  \left\{
    \begin{array}{l}
      \displaystyle
      \dot{x} = y \\
      \dot{y} = z \\
      \dot{z} = -\alpha z + x y^2 -x \:,
    \end{array}
  \right.
  \label{eq:malasoma}
\end{equation}
with chaotic dynamics as solutions when $\alpha \in [2.027;2.08]$. A detailed analysis of the topological properties of the attractors that can be produced by this system has been proposed in \cite{Rosalie_2013,Rosalie_2015}.
For instance, Fig.~\ref{fig:matrix1} summarizes some steps of the topological characterization (Poincaré section and template) of a chaotic attractor when $\alpha = 2.027$.
In this article, we are considering only attractors bounded by genus--1 torus such as Rössler attractors \cite{R_ssler_1976} or Malasoma attractors \cite{Malasoma_2000} (Fig.~\ref{fig:matrix1}a); it does not work for more complex attractors such as Lorenz attractors \cite{Lorenz_1963} bounded by a genus--3 torus. 

\begin{figure}[tbp]
  \centering
  \begin{tabular}{cccccc}
    \multicolumn{2}{c}{Convention} &
    \multicolumn{2}{c}{Torsions} &
    \multicolumn{2}{c}{Permutations} \\
    \includegraphics[height=3em]{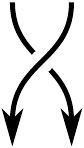} &
    \includegraphics[height=3em]{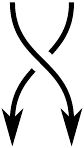} &
    \includegraphics[height=3em]{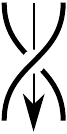} &
    \includegraphics[height=3em]{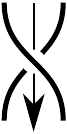} &
    \includegraphics[height=3em]{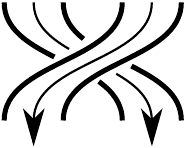} &
    \includegraphics[height=3em]{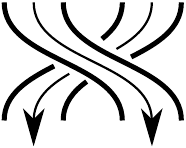} \\
    $+1$ &
    $-1$ &
    positive &
    negative &
    positive &
    negative
  \end{tabular}
  \caption{
    Convention of representing oriented crossings. The permutation
    between two branches is positive if the crossing generated is equal to
    $+1$, otherwise it is negative. We use the same
    convention for torsions.
  }
  \label{fig:convention}
\end{figure}

A \emph{template} is a compact branched two-manifold with boundary and smooth expansive semiflow built locally from two types of charts: joining and splitting \cite{Ghrist_1997}.
It is a figure that represents the topological structure of a chaotic attractor. 
Since the 1990s there have been two different ways to represent templates with linking matrices that are still used today, as one can see in the recent paper of Gilmore and Rosalie \cite{Gilmore_2016}, where algorithms are given to switch from one representation to the other. Hereinafter, the representation first given by Melvin and Tufillaro \cite{Melvin_1991} is considered. This representation only requires a linking matrix, and gives a standard representation at the end, where at the bottom of the template the strips are ordered from the back-most on the left to the front-most on the right. This is the representation used for the template shown in Fig.~\ref{fig:matrix1}. We also use the orientation convention defined by Tufillaro \textit{et al.} \cite{Melvin_1991,Mindlin_1990} (Fig.~\ref{fig:convention}).

A linking matrix is a matrix that details the number and the direction of crossings in a template. As illustrated in Fig.~\ref{fig:convention}, a torsion is a twist of a branch with itself and a permutation is an exchange of position of two branches. Furthermore, the torsions and permutations can be either positive or negative as defined by the orientation convention shown in Fig~\ref{fig:convention}. The linking matrix $M$ corresponding to Fig.~\ref{fig:matrix1} is given by (\ref{eq:matrix1}).
\begin{equation}
 M=
 \left[\begin{matrix}
1 & 0 & 0 & 0 \\
0 & 0 & -1 & -1 \\
0 & -1 & -1 & -1 \\
0 & -1 & -1 & 0
\end{matrix}
\right\rrbracket
\label{eq:matrix1}
\end{equation}

The diagonal elements in the linking matrix correspond to the torsions. As an example, consider matrix $M$. The element $M_{1,1}=1$ represents the number of torsions of branch one of the template from Fig.~\ref{fig:matrix1}. This branch performs exactly one single positive torsion as indicated by the matrix $M$. The non-diagonal elements correspond to the number of permutations between the different branches. As an example, $M_{2,4}=-1$ means that branches two and four perform a negative permutation which is depicted by the crossing of the orange and red branch in Fig.~\ref{fig:matrix1}. It is sufficient to consider the part of the matrix above the diagonal, as it is symmetric.

The linking matrix $M$ is unique but the corresponding template can be drawn in various ways.
Some representations can be longer than others. This is why our goal is to find the most concise template.
This means that we aim to maximize the number of permutations per level of the template. There might be however several templates with minimum size. In this work we only consider the first template of minimum size generated by the algorithm.

An important remark is that not every matrix corresponds to a valid template of a chaotic attractor. As a chaotic attractor is a solution of a deterministic process and the linking matrix represents it, such a matrix needs to fulfill certain criteria.
We will describe the tool we created to verify the validity of a linking matrix, to solve the underlying scheduling problem to find the order of the permutations and to determine the most concise representation of a template. Finally, the tool also renders the solution found.

\section{Related Work}
\label{sec:related work}
The visualization of a template has been addressed in Chap.~5 Sec.~5 of \cite{Tufillaro_1992} and, according to our best knowledge, the validation of a linking matrix has never been addressed.
Usually, this has been done manually by each author. The only comparable project we found is a Mathematica code written by N. B. Tufillaro \textit{et al.} \cite{Tufillaro_1992}, which draws templates. Extensive details are available in the Chap.~5 of \cite{Tufillaro_1992}. It has been used recently in papers written by Barrio \textit{et al.} \cite{Barrio_2009,Barrio_2012,Barrio_2015}. This implementation, however, only works on older versions of Mathematica. Furthermore, one has to specify as input an explicit order of crossings, which means that it does not find them automatically from a linking matrix, unlike the algorithm presented in this paper. This Mathematica code does not provide a validity verification either, it is purely a tool for drawing ``clean'' templates.

To the best of our knowledge, such a tool has never been proposed and could be beneficial for the scientific community, as it is not always easy to see whether a matrix is valid or not. Indeed there have been publications with invalid matrices that our tool would have marked as such \cite{Mindlin_1990}. Some other papers have presented quite unattractive drawings of templates (eg. Fig.~4 of \cite{Anastassiou_2008}) and we feel that our tool would provide researchers with an easy and rapid way to solve this problem. Moreover, it can also be used by the community as a tool for building a linking matrix from the linking number numerically obtained during the topological characterization method for attractors bounded by a genus--1 torus (see \cite{Gilmore_1998,Rosalie_2016} for details).

\section{Linking Matrix and Template of a Chaotic Attractor}
\label{sec:linkin matrix and template of CA}
In this section, we are going to discuss the approach we developed in order to check the validity of a given linking matrix, to find a corresponding template of minimal height as well as to visualize it.
Firstly in section \ref{validation of LM}, we will describe the different validation steps which we are applying on a matrix and justify their necessity. Secondly, section~\ref{tree-like} explains the tree construction we use in order to minimize the height of the resulting template and the methods we apply for the visualization of the template. 

\begin{algorithm}
\caption{Drawing of the template of a linking matrix.}
\label{alg:DWG}
\begin{algorithmic}[1]
\State $\text{verify correct matrix input form}$
\State $\text{verify continuitiy constraints of matrix}$
\State $\text{verify determinism constraints of matrix}$
\If {passed all verification steps}:
\State $\text{construct tree}$
\State $\text{find shortest path in tree}$
\State $\text{draw template}$
\EndIf
\end{algorithmic}
\end{algorithm}

\subsection{Validation of a Linking Matrix}
\label{validation of LM}

A linking matrix is a topological representation of a chaotic attractor, hence it needs to satisfy certain constraints linked to the attractor. Essentially, a template consists of strips that are stretched, twisted, folded and glued at the bottom over and over again after a clockwise rotation. We remind that we are only considering templates of attractors bounded by a genus--1 torus.

In order to visualize this, one can imagine having a sheet of paper split into several strips. The behavior of those strips is given by the elements of the matrix. If one can deform the paper in such a way that the paper respects the constraints given by the matrix without having to tear it apart, then the matrix corresponds to a valid template. If tears are unavoidable, no valid template exists. If there is a tearing mechanism in the attractor, we are out of the scope because this means that the attractor is at least bounded by a genus--2 torus.

\subsubsection*{Validation steps}

The steps below evaluate whether or not a linking matrix is valid, i.e., if it corresponds to a chaotic attractor. 

First of all, we need to verify that a matrix is of the right form. A valid linking matrix, by definition, has a certain construction. It is square, symmetric and has integers as values \cite{Melvin_1991}.

The next three validation steps are constraints on the continuity of the template. Going back to the sheet of paper example, these constraints guarantee that no tears occur.
The first of these constraints is linked to the diagonal elements of the matrix. These elements have to respect the condition which dictates that they have to differ by exactly one from their diagonal neighbors. Violating this constraint would result in a discontinuous template. 
Similar to the diagonal constraint, a linking matrix needs to satisfy the condition which states that an arbitrary value in the matrix cannot differ from the values of all of its neighbors by more than one.
Finally, the last continuity constraint is based on the order of the elements on the bottom of the template. From a linking matrix, one needs to be able to obtain a valid order for the template. The order is an array which defines the position of the branches at the bottom of the template after performing the crossings. We obtain this order from the matrix by applying a simple algorithm described in \cite{Melvin_1991}. A valid ordering array contains all branch indexes exactly once. An index being present twice would mean that two branches would end up at the same end position, which is impossible without a tear and therefore would result in an invalid template.

The last two verification steps are linked to the determinism of a chaotic attractor. As stated earlier, chaotic attractors are solutions of dynamic deterministic systems, meaning that from any starting point there is a unique image and no choice is possible. As the template is a topological representation of a chaotic attractor, it also needs to respect its intrinsic properties like determinism. 
The first of those two verifications consists in checking whether the linking matrix has  $2\times2$ sub-matrices located on its diagonal that are not valid. Up to addition of a global torsion (see \cite{Rosalie_2013} for details) there are two $2\times2$ matrices that are not valid, namely B and C: 
\begin{equation}
\left\{ \label{eq:2x2matrix}
 B =
 \left[\begin{matrix}
-1 & \ 0 \\
\ 0 & \ 0
\end{matrix}
\right\rrbracket ,
\	 C =
 \left[\begin{matrix}
0 & \ \ 0 \\
0 & -1
\end{matrix}
\right\rrbracket  ,
\	 C+1=
 \left[\begin{matrix}
1 & 1 \\
1 & 0
\end{matrix}
\right\rrbracket
, \dots
\right\}\;.
\end{equation}
The set \eqref{eq:2x2matrix} corresponds to matrices that are associated to discontinuous templates. 
If the matrix has such a sub-matrix on its diagonal, this means that it presents a choice opportunity at some point and violates the determinism condition. Therefore, it is not valid. 

\begin{figure}[tb]
\centering
\includegraphics[scale=0.3]{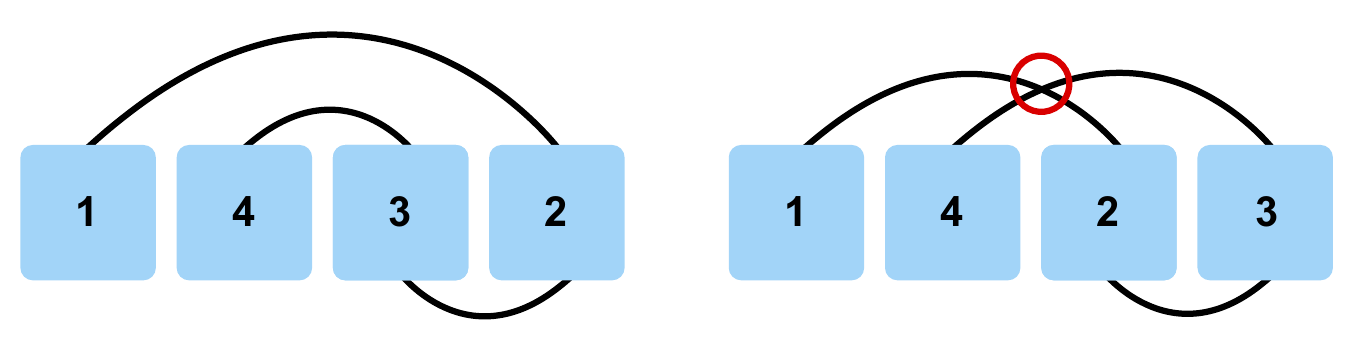}
\caption{Planarity check of matrices \eqref{eq:matrix1} (left) and \eqref{eq:fail_planarity} (right).}
\label{fig:planarity}
\end{figure}

Finally, in the second step, which we call planarity check, we verify the order of the end positions of the template. The idea is to take the final positions of the branches at the end of the template, and connect them with arcs in a certain way. Start with 1, and connect it to 2 over the list. Then connect 2 to 3 below the list, 3 to 4 over, and so on. If the arcs cannot be drawn without intersecting, then the matrix is invalid. This is illustrated by Fig.~\ref{fig:planarity}, where the left part of the figure corresponds to this verification of the matrix \eqref{eq:matrix1}, and has no intersections. The right side on the other hand corresponding to matrix N \eqref{eq:fail_planarity} does not pass the test.
\begin{equation}
\label{eq:fail_planarity}
N =
\left[\begin{matrix}
0 & \ 0 & \  0 & \ 0 \\
0 & \ 1 & \ 0 & -1 \\
0 & \ 0 & \ 0 & -1 \\
0 & -1 & -1 & -1 \\
\end{matrix}
\right\rrbracket
\end{equation}

If this planarity condition was not verified and there was an intersection, the system would have a choice when arriving at this intersection, which would violate the determinism assumption. Therefore, a matrix that does not satisfy this condition cannot correspond to a valid template.


\subsubsection*{Order of the validation steps}

The order of the different validation steps is defined in the way described above, we start checking the most general constraints, and then check the most specific ones (Alg.~\ref{alg:DWG}). For example, if a matrix is not square matrix, there is no need to verify specific constraints like the diagonal constraint, as the matrix is not even a linking matrix by definition. The same idea applies to the other criteria.


In literature, there have been publications with invalid matrices that our procedure would have labeled as such. One example would be the first $4\times 4$ linking matrix in~\cite{Mindlin_1990}, which gives the matrix with the following diagonal elements: $6$, $5$, $5$ and $4$. This matrix would not have passed the validation step which dictates that all elements on the diagonal of a matrix have to differ by one from their diagonal neighbors. 

\begin{equation}
\label{eq:fail_2}
K =
 \left[\begin{matrix}
3 & 2 &  2 & 3 \\
2 & 2 & 2 & 3 \\
2 & 2 &  3 & 4 \\
3 & 3 & 4 & 4 \\
\end{matrix}
\right\rrbracket
\end{equation}
For the matrix K \eqref{eq:fail_2} the ordering validation step fails because the ordering at the end is given by the array $[2,3,3,2]$, meaning that both strips one and four are on position two and strips two and three are on position three. As this is a problem for continuity, this matrix would not pass the order test.
This illustrates that a tool to validate a matrix would facilitate the analysis of linking matrices, as it is not always easy to see whether a matrix is valid or not.
A complete example of the validation process can be found in the appendices of the extended version \cite{Olszewski_2018}. 

\subsection{Visualization of a Template}
\subsubsection{Tree Construction}
\label{tree-like}
After having verified the validity of a linking matrix, the next step is to generate a visualization of a template with minimal height from a given linking matrix. In order to determine the minimal height of a template, one has to optimize the scheduling of all the crossings between the different branches.
For this purpose, we developed an approach where we take as input a valid linking matrix and make use of its permutations to generate a tree graph using a breadth first approach, meaning that we build it level by level. 

%
%
\begin{algorithm}[tbp]
\caption{Tree construction}
\label{alg:tree}
\begin{algorithmic}[1]
	\If {$validMatrix(matrix)$}
    \State $init = Node(permutationList, order, father=None)$ ;
    \State $finalOrder = getFinalOrder(matrix)$
    \State $queue =[init]$ ;
    \While{$queue \neq \varnothing$}
   	\State $node = queue[0]$ ;
    \State $queue = queue[1:]$ ;
    \State $toExecute = permutationList\ \cap \ allNeighborCombinations(node.order)$ ;
    \If {$toExecute = \varnothing$ and $node.order = finalOrder$}
    \State $setLeaf(node)$ ;
    \State $break$ ;
    \EndIf
    \For{$p$ in $toExecute$}
    \State $newNode =$
    \State $Node(updatedPermutationList(p), updatedOrder(p), father=node)$ ;
    \State $queue.append(newNode)$ ;
    \EndFor
    \EndWhile
	\EndIf
\end{algorithmic}
\end{algorithm}

To do this, we follow Alg.~\ref{alg:tree}. We derive the initial order from the matrix which represents the root of the tree as a first step. Furthermore, we also retrieve the list of performable permutations between the branches. Beginning at the root, we simulate the permutations and generate additional nodes which are annotated with an updated order and then added to the tree. For each node created, the list of permutations yet to be performed will differ. Eventually, a node representing a leaf with an empty permutation list and a valid final order will be generated. At this point, the computation of the tree is stopped. By traversing the tree from the root to that leaf, we get the sequence of permutations to execute in order to obtain a template of minimal height. To illustrate this procedure, consider the following $4 \times 4$ matrix A \eqref{eq:matrix_4x4}.
 
 \begin{equation}
 A = 
 \left[\begin{matrix}
-1 & -1 & -1 & -1 \\
-1 & \ 0 & \ 0 & \ 0 \\
-1 & \ 0 & \ 1 & \ 1 \\
-1 & \ 0 & \ 1 & \ 2
\end{matrix}
\right\rrbracket
\label{eq:matrix_4x4}
\end{equation}

\begin{figure}[tb]
\centering
\includegraphics[scale=0.23]{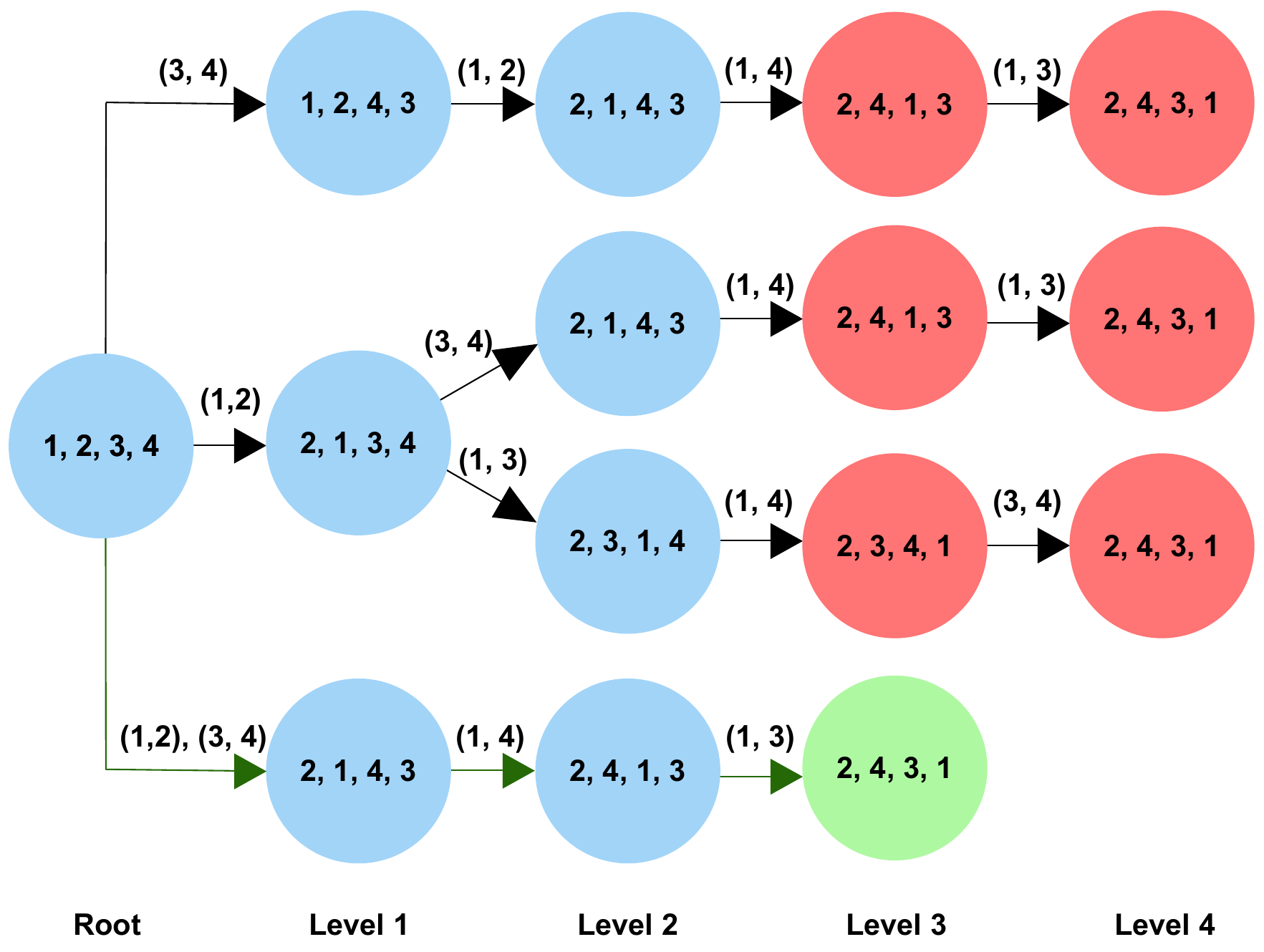}
\caption{Final and complete tree for matrix A from~(\ref{eq:matrix_4x4}) including the root and the child nodes generated per level. Each node represents the updated order after each permutation described by the incoming edge.}
\label{fig:finaltree}
\end{figure}

From this matrix, we get an initial order where the branches are numbered beginning from $1$ to $4$. To retrieve the set of permutations to perform, we have to consider the non-diagonal elements of the matrix. For example, the branch with the label $1$, has to perform a negative permutation with the branches $2$, $3$ and $4$. There is also a positive permutation between branch $3$ and $4$. So, we obtain the following list of permutations which needs to be executed $[(1,2), (1,3), (1,4), (3,4)]$.

To find the permutations which can be performed at this stage, we need to consider our initial order from which we can derive which branches are direct neighbors. For instance, we obtain the following list of neighbor pairs $[(1,2), (2,3), (3,4)]$. By taking the intersection of the neighbor list and the set of permutations to perform, we obtain a set of permutation which are possible to process during the initial stage. By doing so, we can permute branch $1$ and $2$ or $3$ and $4$. However, we could also perform both permutations in parallel as performing one of them does not prohibit the other one. As illustrated on top of Fig.~\ref{fig:finaltree}, we see the root labeled with the initial order of the branches. After the first set of permutations have been performed, different child nodes are created at level $1$. The corresponding order of each child node is obtained by switching the positions of the permuted branches in the initial order of the root.

From the new order of each child node, we try to find a new permutation to perform by defining the neighbor pairs. We then recompute the possible permutations for this iteration. Each iteration will add one or more children to tree and this process is repeated until all permutations have been performed or no new permutation can be computed. However, a node which can no longer perform a permutation while there are still some permutations in the set left to be executed, is not considered valid. 

Fig.~\ref{fig:finaltree} also shows the final tree after all permutations have been performed. The green arrows leading to the green colored leaf denote the shortest path where the labels show the order of execution of the permutations to get to the final order of the template. This will result in a template of shortest possible height. There are also three other possible solutions but they will not reduce the height of the template to a minimum as they perform one additional permutation. However, we stop the computation of building the tree after encountering the first valid leaf, so the red nodes will never be computed. The breadth-first construction of the tree guarantees that the first found solution is the shortest one.

\subsubsection{Drawing of the Template}
Finally, after verification of the linking matrix and after having found the shortest path in the tree corresponding to the most concise order of crossings, we can now draw the template. To draw the templates as scalable vector graphics, we used python's \texttt{swgwrite} module \cite{svgwrite}.

In order to draw both torsions and permutations, we use a cubic B\'ezier curve as shape. To illustrate how we use it, consider two points $(x_1,y_1)$ and $(x_2,y_2)$ and suppose we want to draw this B\'ezier curve between them, in the same shape as those used in the permutations and torsions. The starting point is given by $(x_1,y_1)$ and we will give the rest of the points relative to this starting point. The relative end point is then given by $(x_2-x_1,y_2-y_1)$ and the two relative control points by $(x_1, (y_2-y_1)/2)$ and $(x_2-x_1, (y_2-y_1)/2)$. So the control points are always halfway in height between the two points and straight above respectively below them.

\begin{figure}[tbp]
\setlength{\tabcolsep}{12pt}
  \centering
  \begin{tabular}{ccccc}
    \includegraphics[height=2.5em]{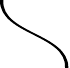} &
    \includegraphics[height=2.5em]{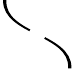} &
    \includegraphics[height=2.5em]{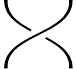} &
    \includegraphics[height=2.5em]{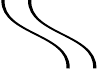} &
    \includegraphics[height=2.5em]{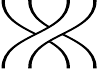} \\
    (a) &
    (b) &
    (c) &
    (d) &
    (e)
  \end{tabular}
  \caption{
An illustration of a positive torsion (a--c) and a positive permutation (d--e) drawing process.}
\label{fig:torsion}
\end{figure}

To draw a torsion we first draw one B\'ezier curve, then add a small white circle in the middle of this curve to \textit{erase} this part. Finally we draw the other B\'ezier curve. This procedure is illustrated in Fig.~\ref{fig:torsion}(a--c).
Permutations are drawn in a similar way. The sign of the permutation defines which of the two branches is drawn first, then when the other one is drawn it covers it up as it comes on top of the other one (Fig.~\ref{fig:torsion}(d--e)).

\begin{figure}[tb]
\centering
\includegraphics[scale=0.20]{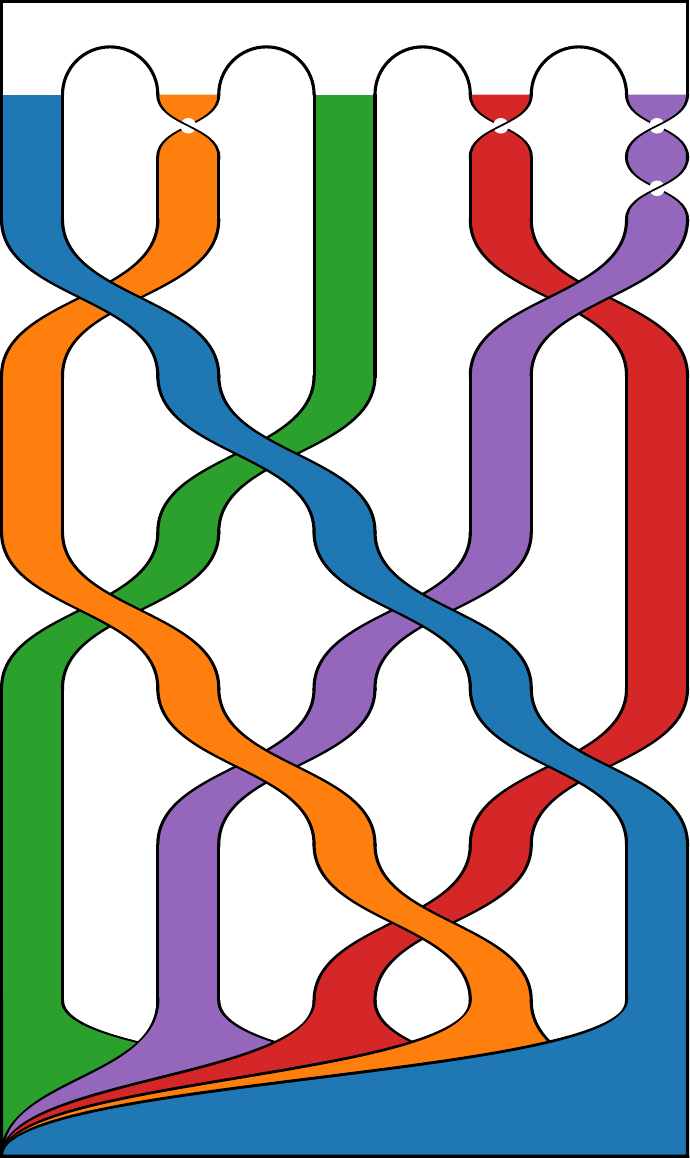}
\caption{Template of one linking matrix with five branches and eight permutations.}
\label{fig:mat5_11}
\end{figure}

We start by considering the torsions of the matrix and draw all of them. Then we move on to the permutations. They are given by the sequence of edges forming the shortest path of the tree generated by the input matrix. We then draw the rest of the template by levels. At each level, every strip can do one of three actions: do a straight transition, permute left or permute right. The shortest path tells us which two strips should permute. Given this information, it is easy to calculate the coordinates at the next level of each strip and apply the correct transition (Fig.~\ref{fig:mat5_11}).

\section{Performance Evaluation}
\label{sec:results}

\begin{figure}[th]
  \centering
    \includegraphics[width=.7\textwidth]{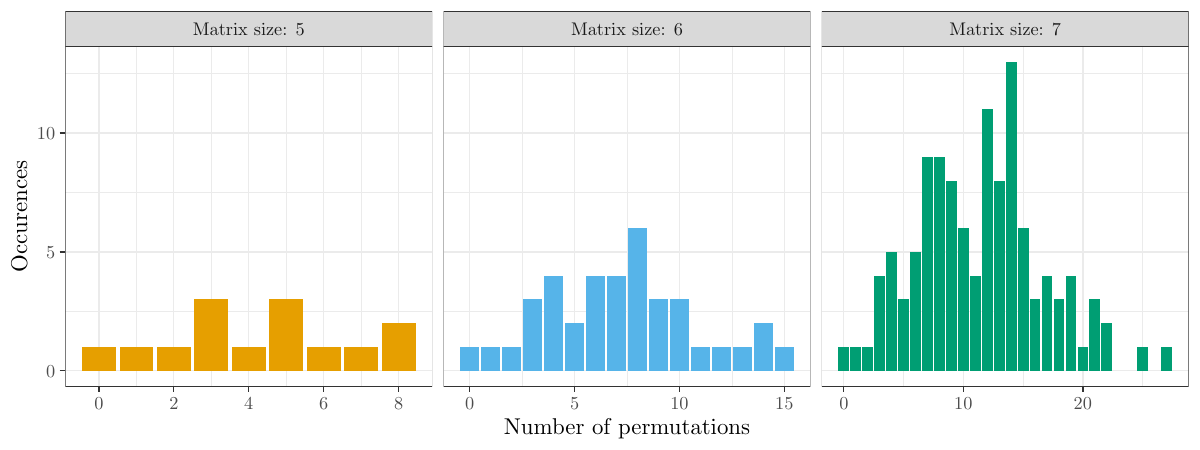} 
  \caption{
  Distribution of the number of elementary matrices with respect to the number of permutations to process. There are 14 (resp.\@ 38 and 116) matrices of size 5 (resp.\@ 6 and 7).
  }
\label{fig:permutations_distributions}
\end{figure}

An elementary matrix is a unique linking matrix describing a chaotic mechanism without additional torsions or symmetry properties \cite{Rosalie_2017}.
Given an input size, Rosalie describes in this article a method to generate all possible elementary linking matrices of such size.
We used this method to obtain the 14, 38 and 116 possible elementary matrices with resp.\@ five, six and seven branches (resp.\@ $5 \times 5$, $6 \times 6$ and $7 \times 7$ linking matrices).
Fig.\@~\ref{fig:permutations_distributions} depicts for each matrix size the distribution of the elementary matrices with respect to the number of permutations to process.

\begin{figure}[b!]
  \centering
    \includegraphics[width=.9\textwidth]{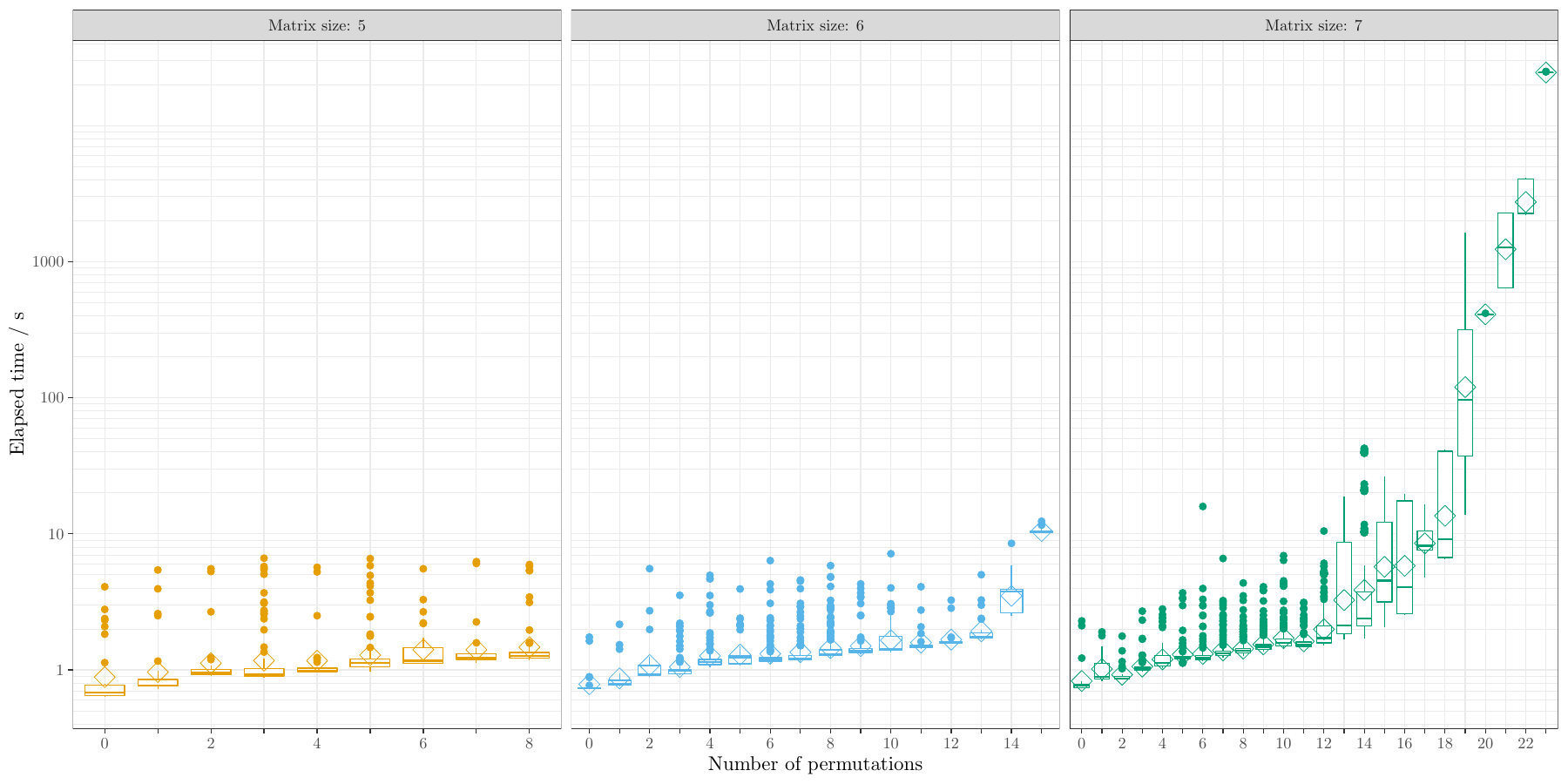} 
  \caption{
  	Elapsed time depending on the number of permutations for the  matrices depending on their size. The diamond represents the average value.
  }
\label{fig:elapsed-time_vs_nb-permutations_detailed-plot}
\end{figure}

\begin{figure}[t!]
  \centering
    \includegraphics[width=.7\textwidth]{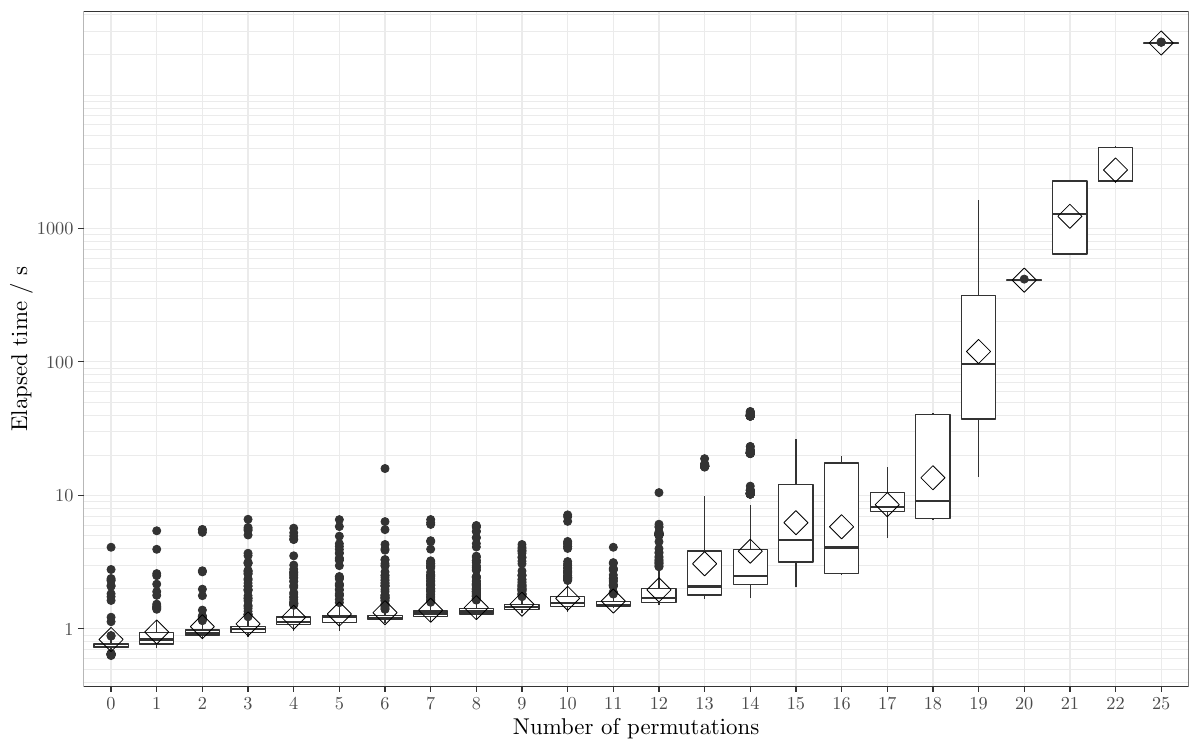}
  \caption{
  	Elapsed time depending on the number of permutations for the 167 matrices. The diamond represents the average value.
  }
\label{fig:elapsed-time_vs_nb-permutations_grouped-plot}
\end{figure}

The experiments were conducted on a server with an Intel Xeon X7560 processor with a clock speed of 2.27~GHz, and 1024~GB of RAM.
Even though this server is not the fastest available, it is the only one fulfilling the memory requirements (instances required between 25~MB and 400~GB of memory).
For the sake of comparability, all instances have been run on the same machine.
For the complete description of the cluster environment, please refer to \url{https://hpc.uni.lu/systems/chaos/}.
We computed the templates of all the elementary matrices described above.
We ran the experiments with version v0.0.1 of the code.
For each input matrix, we measured 30~times the time elapsed to get the template.
The $7 \times 7$ matrix with 27~permutations ran out of memory and crashed: we removed it from the graphs.
Fig.\@~\ref{fig:elapsed-time_vs_nb-permutations_detailed-plot} and Fig.\@~\ref{fig:elapsed-time_vs_nb-permutations_grouped-plot} depict the elapsed computation time with respect to the number of permutations to process.
As expected, we observe a drastic rise that characterizes a combinatorial explosion in the number of permutations.

\section{Conclusion}
\label{sec:conclusion}

In this paper, we presented a tool which verifies whether a linking matrix corresponds to a topologically valid template. Moreover, our approach computes and draws a template of minimal height corresponding to this linking matrix. This is especially interesting for linking matrices with a higher number of crossings. We believe that this tool could benefit the research community as it eases the process of verifying the validity of a linking matrix, and quickly draws one of its matching templates.

A possible extension of our work could be to represent the generated templates as a 3D model in an automated way. One representation of a 3D template was given by Cross and Gilmore, where they include the torsions as a part of the global modification \cite{Cross_2013}. Another visualization was given by Boulant \textit{et al.} (Fig.~6 of \cite{Boulant_1998}). Such a 3D visualization would allow to be even closer visually to the nature of a chaotic attractor, and thus could provide more intuitive insights.

\subsection*{Acknowledgments}
The experiments presented in this paper were carried out using the HPC facilities of the University of Luxembourg \cite{Varrette_2014} (see \url{https://hpc.uni.lu}).
This work is partially funded by the joint research programme UL/SnT-ILNAS on Digital Trust for Smart~ICT.

\newpage
\bibliographystyle{splncs04}
\bibliography{mybiblio}

\end{document}